\documentclass[12pt]{article} 
\usepackage{amsmath,amsfonts,amssymb,latexsym,cite,bm} 
\usepackage{slashed} 
\usepackage{relsize} 
\usepackage{graphicx} 

\usepackage{prodint}

\newcounter{thMM}
\newcounter{leMM}
\newcounter{deFF}
\newcounter{exMP}
\title{
 \Large\bf Schr\"odinger wave functional in quantum Yang-Mills theory from precanonical quantization 
\\
}
\author{ 
Igor V. Kanatchikov
\\ 
\small\it School of Physics and Astronomy \\ 
\small\it University of St Andrews, St Andrews KY16 9SS, UK \\
\small\it E-mail: {\tt ik25@st-andrews.ac.uk} 
\\ [1ex]
\small\it Quantum Information Center in Gda\'nsk (KCIK),\\ 
\small\it 81-831 Sopot, Poland  
}

\catcode `\@=11
\@addtoreset{equation}{section}

\def\section{\@startsection {section}{1}{\z@}{-3.5ex plus -1ex minus
     -.2ex}{2.3ex plus .2ex}{\normalsize\bf}}
\def\subsection{\@startsection{subsection}{2}{\z@}{-3.25ex plus -1ex minus
 -.2ex}{1.5ex plus .2ex}{\normalsize\bf}}

 \def\thebibliography#1{\section*{References\markboth
  {REFERENCES}{REFERENCES}}\list
  {[\arabic{enumi}]}{\settowidth\labelwidth{[#1]}\leftmargin\labelwidth
  \advance\leftmargin\labelsep
  \usecounter{enumi}}
  \def\newblock{\hskip .11em plus .33em minus -.07em}
  \sloppy
  \sfcode`\.=1000\relax}
 
\catcode `\@=12

\begin{document}

\date{}

\maketitle
\begin{abstract}
\noindent 
{\small A relation between the precanonical quantization of pure Yang-Mills fields 
and the standard functional Schr\"odinger 
representation in the temporal gauge is discussed. It is shown that the 
latter can be obtained  from the former when the ultraviolet parameter $\varkappa$ 
introduced in precanonical quantization goes to infinity. In this limiting case, the 
Schr\"odinger wave functional can be expressed as the trace of the Volterra product 
integral  of Clifford-algebra-valued precanonical wave  functions 
restricted to a 
field configuration,  and the canonical functional derivative Schr\"odinger equation  
together with  
the quantum Gau\ss\ constraint can be derived 
from the Dirac-like precanonical Schr\"odinger equation. 
}

\medskip

\end{abstract}


\newcommand{\beq}{\begin{equation}}
\newcommand{\eeq}{\end{equation}}
\newcommand{\beqa}{\begin{eqnarray}}
\newcommand{\eeqa}{\end{eqnarray}}
\newcommand{\nn}{\nonumber}

\newcommand{\half}{\frac{1}{2}}

\newcommand{\xt}{\tilde{X}}

\newcommand{\uind}[2]{^{#1_1 \, ... \, #1_{#2}} }
\newcommand{\lind}[2]{_{#1_1 \, ... \, #1_{#2}} }
\newcommand{\com}[2]{[#1,#2]_{-}} 
\newcommand{\acom}[2]{[#1,#2]_{+}} 
\newcommand{\compm}[2]{[#1,#2]_{\pm}}

\newcommand{\lie}[1]{\pounds_{#1}}
\newcommand{\co}{\circ}
\newcommand{\sgn}[1]{(-1)^{#1}}
\newcommand{\lbr}[2]{ [ \hspace*{-1.5pt} [ #1 , #2 ] \hspace*{-1.5pt} ] }
\newcommand{\lbrpm}[2]{ [ \hspace*{-1.5pt} [ #1 , #2 ] \hspace*{-1.5pt}
 ]_{\pm} }
\newcommand{\lbrp}[2]{ [ \hspace*{-1.5pt} [ #1 , #2 ] \hspace*{-1.5pt} ]_+ }
\newcommand{\lbrm}[2]{ [ \hspace*{-1.5pt} [ #1 , #2 ] \hspace*{-1.5pt} ]_- }

\newcommand{\pbr}[2]{ \{ \hspace*{-2.2pt} [ #1 , #2\hspace*{1.4 pt} ] 
\hspace*{-2.3pt} \} }
\newcommand{\nbr}[2]{ [ \hspace*{-1.5pt} [ #1 , #2 \hspace*{0.pt} ] 
\hspace*{-1.3pt} ] }

\newcommand{\we}{\wedge}
\newcommand{\nbrpq}[2]{\nbr{\xxi{#1}{1}}{\xxi{#2}{2}}}
\newcommand{\lieni}[2]{$\pounds$${}_{\stackrel{#1}{X}_{#2}}$  }

\newcommand{\rbox}[2]{\raisebox{#1}{#2}}
\newcommand{\xx}[1]{\raisebox{1pt}{$\stackrel{#1}{X}$}}
\newcommand{\xxi}[2]{\raisebox{1pt}{$\stackrel{#1}{X}$$_{#2}$}}
\newcommand{\ff}[1]{\raisebox{1pt}{$\stackrel{#1}{F}$}}
\newcommand{\dd}[1]{\raisebox{1pt}{$\stackrel{#1}{D}$}}
\newcommand{\der}{\partial}
\newcommand{\oo}{$\Omega$}
\newcommand{\Om}{\Omega}
\newcommand{\om}{\omega}
\newcommand{\eps}{\epsilon}
\newcommand{\si}{\sigma}
\newcommand{\Lm}{\bigwedge^*}

\newcommand{\inn}{\hspace*{2pt}\raisebox{-1pt}{\rule{6pt}{.3pt}\hspace*
{0pt}\rule{.3pt}{8pt}\hspace*{3pt}}}
\newcommand{\sro}{Schr\"{o}dinger\ }
\newcommand{\vol}{\omega}
               \newcommand{\dvol}[1]{\der_{#1}\inn \vol}

\newcommand{\bd}{\mbox{\bf d}}
\newcommand{\bder}{\mbox{\bm $\der$}}
\newcommand{\bI}{\mbox{\bm $I$}}

\newcommand{\be}{\beta} 
\newcommand{\ga}{\gamma} 
\newcommand{\de}{\delta} 
\newcommand{\Ga}{\Gamma} 
\newcommand{\gmu}{\gamma^\mu}
\newcommand{\gnu}{\gamma^\nu}
 \newcommand{\ka}{\varkappa} 
 \newcommand{\la}{\lambda}
\newcommand{\hka}{\hbar \kappa}
\newcommand{\al}{\alpha}
\newcommand{\lapl}{\bigtriangleup}
\newcommand{\psib}{\overline{\psi}}
\newcommand{\Psib}{\overline{\Psi}}
\newcommand{\Phib}{\overline{\Phi}}
\newcommand{\derts}{\stackrel{\leftrightarrow}{\der}}
\newcommand{\what}[1]{\widehat{#1}}

\newcommand{\bx}{{\bf x}}
\newcommand{\bk}{{\bf k}}
\newcommand{\bq}{{\bf q}}

\newcommand{\omk}{\omega_{\bf k}} 
\newcommand{\lpl}{\ell}
\newcommand{\zb}{\overline{z}} 

\newcommand{\deltab}{\bar \delta}

\newcommand{\dv}{\mbox{\sf d}}

\newcommand{\deltt}{\bm{\delta}}   

\newcommand{\BPsi}{\mathbf{\Psi}} 
\newcommand{\BPhi}{\mathbf{\Phi}}
\newcommand{\BH}{{\bf H}} 
\newcommand{\BS}{{\bf S}} 
\newcommand{\BN}{{\bf N}}

\newcommand{\rd}{\mathrm{d}}
\newcommand{\ri}{\mathrm{i}}
\newcommand{\Tr}{\mathrm{Tr}} 


\section{Introduction}

The canonical quantization of YM field theory in the 
functional Schr\"odinger picture of QFT 
 \cite{feynm,gaw,rossi,hatf85,jackiw,luscher,kim,hatfield,mansfield,islam,pachos,olej,krug} 
leads (in the temporal gauge $A^a_0(\bx) = 0$) 
to the description of the corresponding 
quantum field in terms of the Schr\"odinger wave functional 
$\BPsi([A^a_i(\bx)],t)$  which satisfies the 
functional derivative Schr\"odinger equation 
\beq \label{fseq}
\ri\hbar \der_t \BPsi = 
 \int \! d\bx\, 
 \left ( - \frac{\hbar^2}{2}\ \frac{\delta}{\delta A^i_a(\bx)}\frac{\delta}{\delta A^i_a(\bx)} 
+ \frac{1}{4} F_a{}^{ij}(\bx) F_a{}^{ij}(\bx)  
 \right) \BPsi 
\eeq
and the Gau\ss\ law constraint on the physical states    
\beq \label{gauss}
\left ( \der_i \frac{\delta}{\delta A^a_i(\bx)} \, + \,  
g C^a{}_{bc} A^b_i (\bx)\frac{\delta}{\delta A^c_i(\bx)} 
\right ) \BPsi = 0 \,   
\eeq
which picks the gauge invariant functionals $\BPsi$. 
 
 Precanonical quantization of YM fields proposed in \cite{my-ym1,my-ehrenfest} 
 leads  to the description 
 in terms of the Clifford-algebra-valued wave function 
$\Psi(A^a_\mu, x^\mu)$ which satisfies 
the partial derivative precanonical   Schr\"odinger equation 
\beq  \label{ym-nse}
\ri\hbar \gamma^\mu\der_\mu \Psi = 
\left(\frac{1}{2} \hbar^2\varkappa \frac{\der^2}{\der A_a^\mu\der A^a_\mu } 
- \frac{1}{2} \ri g\hbar  C^a{}_{bc}A^b_\mu A^c_\nu 
\gamma^\nu \frac{\der}{\der A^a_\mu } \right)\Psi =: \frac1\ka\what{H}\Psi \, 
\eeq
and the constraint 
\beq \label{qconstr}
\what{\pi}{}^{(\nu\mu)}_a \Psi = 
- \ri\hbar \ka
\left( \gamma^\mu \frac{\der}{\der A^a_\nu } + \gamma^\nu \frac{\der}{\der A^a_\mu } \right)\Psi 
\approx 0 \,,
\eeq
which is the quantum counterpart of the antisymmetry of the classical YM field strength 
\beq
F^a_{\mu\nu} := \der_\mu A^a_\nu - \der_\nu A^a_\mu 
+ g C^a{}_{bc} A^b_\mu A^c_\nu \,. 
\eeq
The notations and conventions used throughout this paper follow \cite{my-ym1,my-ehrenfest}.

In precanonical quantization \cite{my-ehrenfest,qs96,bial97,lodz98}, one introduces an ultraviolet parameter $\ka$ of dimension of the inverse spatial volume. One of the appearances of this parameter is in the representation of 
the differential form associated with an infinitesimal spatial volume element $\rd\bx $
 as an element of the Clifford algebra of the corresponding $n$-dimensional pseudo-Euclidean space-time   
$\mathbb{R}^{1,n-1}$ (our signature convention is $+---...$) 
\beq \label{qmap}
\rd\bx :=\rd x^1 \we \rd x^2 \we... \we\rd x^{n-1}  \mapsto \frac{1}{\ka} \gamma_0 \,.
\eeq
In what follows, we  use the standard notation $\gamma_0=\beta$ with $\beta^2=1$ and set $\hbar=1$ for convenience. 

In this paper, we show that the description of quantum YM fields in the 
functional Schr\"odinger picture of QFT 
 obtained from canonical quantization follows 
from the quantum YM theory - obtained by precanonical quantization - 
in the singular limiting case corresponding to the infinite value of the ultraviolet parameter 
$\ka$ introduced by precanonical quantization. 
 
From the point of view of 
 standard QFT, at first glance,  the claim that 
(\ref{ym-nse}) may have something to do with (\ref{fseq}) sounds improbable at least. 
Nevertheless, in \cite{atmp1,atmp2}, improving an earlier consideration in \cite{my-pla}, 
we have already shown that, in the case of interacting scalar fields, the 
 description of quantum fields in the functional Schr\"odinger picture 
 can be derived from the precanonical description in the limiting case corresponding to the 
 inversion of the map (\ref{qmap}),  i.e. when $\frac{1}{\ka}$ is infinitesimal. 
 In this paper, we will show that a similar relation exists between the functional Schr\"odinger   equation and the precanonical Schr\"odinger equation for quantum pure YM field. 
 The insights from \cite{atmp1,atmp2} 
 will be closely followed in the present consideration. 
 
 Intuitively, the relation between the Schr\"odinger wave functional and the precanonical wave function is based on the probabilistic interpretation of the former,  as the probability amplitude of finding a field configuration $A(\bx)$ at some moment of time $t$, 
 and the latter, as the probability amplitude of observing the field value $A$ at the space-time point $x$,   that allows us to expect that the time dependent complex functional probability amplitude $\BPsi([A(\bx)], t)$ is a composition of space-time dependent Clifford-valued probability amplitudes given by the precanonical wave function  $\Psi(A,x)$.  
 
 Let us recall that 
 precanonical quantization \cite{qs96,bial97,lodz98,geom-q} 
 is based on the De Donder-Weyl (DW) Hamiltonian theory \cite{dw} which treats space-time variables on  equal footing. In this formulation,  
 the Poisson brackets are defined on differential forms representing the dynamical variables, 
 that leads to the Poisson-Gerstenhaber algebra structure generalizing the Poisson algebra in the canonical Hamiltonian formalism \cite{geom-q,pg} (see also \cite{joseph,romer,my-dirac}). 
 The result of its quantization (a small Heisenberg-like subalgebra of it) is that both the  operators and wave functions  are Clifford-algebra-valued, and the 
 precanonical Schr\"odinger equation includes the space-time Dirac operator 
 instead of the standard time derivative. 
 One of the striking features of this formulation of quantized fields 
 is that, as a consequence of the evolution of precanonical wave function given by the precanonical Schr\"odinger equation, it allows us to reproduce the classical field equations as the equations of expectation values of operators defined by precanonical quantization, thus generalizing the Ehrenfest theorem \cite{my-ehrenfest}. Moreover, by treating the space-time variables on equal footing,   the precanonical quantization approach provides a very natural and promising framework for  quantization of  gravity \cite{ijtp2001,qg}.

Other recent discussions of gauge fields and gravity from the point of view of 
classical DW Hamiltonian theory and its geometrizations can be found in \cite{marco,bruno,helein1,mex1,mex2,romanroy18,dvey}.  

 Most recently, the consideration of the eigenvalue problem for the 
 DW Hamiltonian operator $\what{H}$ on the right-hand side of (\ref{ym-nse}) has allowed us to 
  estimate the 
 non-vanishing gap in the spectrum of quantum SU(2) pure YM theory on $\mathbb{R}^n$ 
 obtained from the precanonical quantization
 \cite{my-massgap}. This paper, by establishing a connection of precanonical approach 
 with the standard formulation of quantum YM theory in the functional Schr\"odinger  representation, can be viewed as an additional substantiation of the identification of the gap in the spectrum of precanonical DW Hamiltonian operator with the mass gap in the energy spectrum of the canonical Hamiltonian operator.

\section{From precanonical wave function to the Schr\"odinger wave functional} 

Our task is to investigate a connection between the functional Schr\"odinger 
representation of quantum YM theory and precanonical quantization. 
In the case of interacting scalar fields this connection was established in 
Refs. \cite{atmp1,atmp2} which refine an earlier treatment in \cite{my-pla}. 
 Similarly, our current treatment can be 
seen as an improvement of our earlier discussion in \cite{my-ym1} which was based on \cite{my-pla}. 

A possible existence of a relation between the functional Schr\"odinger picture and 
the precanonical descripton  implies that 
 the  Schr\"odinger wave functional $\BPsi ([A(\bx)],t)$ 
 can be expressed as a functional of precanonical wave functions 
restricted to a specific field configuration $\Sigma$: $A=A(\bx)$ at time $t$: 
\beq \label{psib}
\BPsi([A(\bx)],t) = \BPsi ([\Psi_\Sigma (\bx,t), A (\bx)]) \,,  
\eeq
where we have denoted the restriction of precanonical wave function $\Psi(A,x)$  to $\Sigma$ as 
$\Psi_\Sigma (\bx,t) := \Psi (A=A(\bx), \bx, t)$. 
Then the time derivative of $\BPsi$ is obtained by the chain rule differentiation 
\beq \label{dtpsi0}
 \ri\der_t \BPsi = 
 {\Tr} \int\! d\bx\, \left \{ 
 \frac{\delta \BPsi }{\delta\Psi^T_\Sigma(\bx, t)} 
\ri\der_t \Psi_\Sigma (\bx, t)  
\right \} \,, 
 \eeq
where $\Psi^T$ is the transpose matrix of $\Psi$. In what follows, for brevity, we will 
denote $\Psi_\Sigma (\bx, t)$  simply as $\Psi_\Sigma (\bx)$ or even $\Psi_\Sigma $ when  appropriate.  

The first and the second 
 {\em total} 
 variational 
 derivatives of  $\BPsi$ with respect to $A^a_\mu(\bx)$ 
 (denoted as $\frac{\deltt  }{\deltt A^a_\mu(\bx)}$) 
 take the form  
\beqa \label{delta-bpsi1} 
 \frac{\deltt \BPsi }{\deltt A^a_\mu(\bx)}  
&=&   
\frac{\deltab \BPsi }{\deltab A^a_\mu(\bx)}  + 
\Tr\left \{ 
 \frac{\delta \BPsi }{\delta\Psi^T_\Sigma(\bx)} 
\der_{A^a_\mu} \Psi_\Sigma (\bx )  
\right \} 
,
\\
 && \nn \\
  && \nn \\
\frac{\deltt{}^2 \BPsi }{\deltt A^a_\mu(\bx)\deltt A^a{}^\mu(\bx)}
&=& 
\frac{\deltab^2 \BPsi }{\deltab A^a_\nu (\bx)\deltab A^a{}^\nu  (\bx)}  + 
\Tr\left \{ 
 \frac{\delta \BPsi }{\delta\Psi^T_\Sigma(\bx )} 
~\delta(\mbox{\bf 0})\der_{A^a_\mu}\der_{A^a{}^\mu} \Psi_\Sigma (\bx)  
\right \}
  \nn \\ 
  && \nn \\
&+&  \Tr \, \Tr \left \{ 
 \frac{\delta^2 \BPsi}{\delta \Psi^T_\Sigma(\bx)\otimes\delta\Psi^T_\Sigma(\bx)} 
~\der_{A^a_\mu} \Psi_\Sigma (\bx)  
\otimes  \der_{A^a{}^\mu} \Psi_\Sigma  (\bx) \right\}
\label{delta-bpsi2} \\
&& \nn \\
&+& 2 \Tr\left\{  \frac{\delta \deltab \BPsi }{\delta \Psi^T_\Sigma(\bx)\deltab A^{a\nu}(\bx)} 
\der_{A^a_\nu} \Psi_\Sigma(\bx)
\right \} ,
 \nn
\eeqa 
where $\frac{\deltab  }{\deltab A^a_\mu(\bx)}$ denotes 
the {\em partial} functional derivative   with respect to $A^a_\mu(\bx)$, $\der_{A^a_\mu}$ is the same as $\frac{\der}{\der {A^a_\mu}}$  and 
$\delta(\mbox{\bf 0})$ is the  $(n-1)$-dimensional delta-function at $\bx=0$ 
 which results from the variational differentiation 
of the function $\Psi_\Sigma (\bx)$ with respect to itself at the same spatial point. 
Here and in what follows, when writing $\delta(\mbox{\bf 0})$, 
 we imply that a 
proper regularization like the split point  or a lattice one has been applied in order 
to make sense of this singular expression. This is the regularization which is usually implied when the second variational derivative is used in the functional Schr\"odinger equation in QFT.

The time derivative $\ri\der_t \Psi_\Sigma (\bx, t)$ is determined by the restriction of 
precanonical Schr\"o\-dinger equation 
(\ref{ym-nse}) 
to  $\Sigma$, which takes the form 
\beqa \label{psisigmaeqn}
\ri\der_t \Psi_\Sigma = -\ri\alpha^i \left ( \frac{d}{dx^i} 
- \der_{i} A{}^a_{\mu} (\bx) \frac{\der}{\der A^a_\mu } \right ) \Psi_\Sigma 
+ \beta \frac1\ka \what{H}_\Sigma \Psi_\Sigma~, 
\eeqa 
where  $\frac{d}{dx^i} $ is the total derivative along $\Sigma$: 
\beq \label{ttl}
\frac{d}{dx^i}:= \der_i + \der_i A^a_\mu(\bx)  \frac{\der}{\der A^a_\mu } 
+ \der_i A^a_{\mu,k}(\bx) \frac{\der}{\der A^a_{\mu,k} } +... , 
\eeq
$A^a_{\mu,k}$ denote the fiber coordinates of the 
first-jet bundle of the bundle of YM potentials $A^a_\mu$ over space-time 
(c.f. \cite{saunders}) 
and $\what{H}_\Sigma$ is the restriction to $\Sigma$ of the operator 
  on the right-hand side of 
(\ref{ym-nse}): 
\beq \label{dwhop}
\frac1\ka\what{H} = 
 \frac{1}{2} \varkappa \frac{\der^2}{\der A_a^\mu\der A^a_\mu } 
- \frac{1}{2}\ri g   C^a{}_{bc}A^b_\mu A^c_\nu 
\gamma^\nu \frac{\der}{\der A^a_\mu } \,,
\eeq 
which is called the De Donder-Weyl Hamiltonian operator (of YM field) \cite{my-ym1,my-ehrenfest}. 
As $\what{H}$ contains no spatial (horizontal) derivatives, its restriction $\what{H}_\Sigma = \what{H}$. 
 Similarly, the restricted wave function $\Psi_\Sigma (\bx)$ is also required to obey 
the restriction of the constraints (\ref{qconstr}) to $\Sigma$:
 \beq \label{qconstr2}
 \hat{\pi}{}_a^{(\mu\nu)}\Psi_\Sigma (\bx) \approx 0 \Leftrightarrow 
 (\gamma^\mu\der_{A^a_\nu} + \gamma^\nu\der_{A^a_\mu})\Psi_\Sigma (\bx)\approx  0  \,. 
\eeq 
 Let us note here the recent discussion of the classical counterpart of this constraint 
 in \cite{mex1}.

Now, by  substituting (\ref{psisigmaeqn}) into (\ref{dtpsi0}) and using (\ref{dwhop}), 
we obtain 
\beqa \label{dtpsi}
\ri\der_t \BPsi &=& \int\!d\bx\ \mathtt{Tr} \Big\{ \frac{\delta \BPsi}{\delta \Psi^T_\Sigma (\bx)}
 \Big( \underbrace{-\ri\alpha^i \frac{d}{dx^i}}_{I} 
+ \underbrace{\ri\alpha^i\der_{i} A{}^a_{j} (\bx) \frac{\der}{\der A^a_j}}_{II} 
+ \underbrace{\ri\alpha^i\der_{i} A{}^a_{0} (\bx) \frac{\der}{\der A^a_0 }}_{III} 
\nn \\ 
&& \underbrace{-\frac12 \ka \beta \frac{\der}{\der{A^a_i}} \frac{\der}{\der{A^a_i}}}_{IV} 
+ \underbrace{\frac12 \ka \beta \frac{\der}{\der{A^a_0}} \frac{\der}{\der{A^a_0}}}_{V} 
\underbrace{-\frac12 \ri g\beta C^a_{bc}A^b_iA^c_j\gamma^j \frac{\der}{\der{A^a_i}}}_{VI} %
 \nn \\
&& 
\underbrace{-\frac12 \ri g\beta C^a_{bc}A^b_iA^c_0\gamma^0\frac{\der}{\der{A^a_i}}}_{VII} %
\underbrace{-\frac12 \ri g\beta C^a_{bc}A^b_0A^c_j\gamma^j \frac{\der}{\der {A^a_0}}}_{VIII} %
 \Big)  \Psi_\Sigma(\bx) \Big\} . 
\eeqa
Let us see if this equation can reproduce the functional derivative 
Schr\"odinger equation (\ref{fseq}) in some sense. 

\bigskip

Due to the constraint $\hat{\pi}{}_a^{(ij)}\Psi_\Sigma \approx 0$ in (\ref{qconstr2})  
the term $(II)$ takes the form 
\beq
\mathit{II}:\quad \ri\alpha^i\der_{[i} A{}^a_{j]} (\bx) \frac{\der}{\der A^a_j }
\Psi_{\Sigma}(\bx)
.
\eeq 
By combining it with the   term $(VI)$, we obtain 
\beq 
\frac\ri2\beta\gamma^iF^a_{ij}\der_{A^a_j} \Psi_{\Sigma}(\bx) .
\eeq 
Then, using the constraint $\hat{\pi}{}_a^{(i0)}\Psi_\Sigma \approx 0$,  
we transform the latter to the form 
\beq \label{term2+6}
\mathit{II+VI}:\quad 
 \frac\ri2 \gamma^{ij}F^a_{ij}\der_{A^a_0} \Psi_{\Sigma}(\bx) . 
\eeq
 
Further, by comparing the terms $(IV)$ and $(V)$ in (\ref{dtpsi})  with the expression for the 
total second variational derivative of $\BPsi$ in (\ref{delta-bpsi2}),  
we conclude that the latter can  be reproduced 
 if $\beta\ka$ is replaced by the (regularized value of the) 
 $(n-1)$-dimensional Dirac delta function at equal spatial points 
 $\delta(\mathbf{0})$:  
\beq \label{722}
\beta\ka \mapsto \delta(\mathbf{0}) . 
\eeq 
In this case, using (\ref{delta-bpsi2}) the terms $(IV)$ and $(V)$ lead to 
\beqa 
  \Tr\left \{\frac{\delta \BPsi }{\delta\Psi^T_\Sigma(\bx, t)} 
 \beta\ka\, \der_{A^a_\mu}\der_{A_a^\mu} \Psi_\Sigma (\bx) \right \}
 &\mapsto &
\frac{\deltt{}^2 \BPsi }{\deltt A_a^\mu(\bx)\deltt A^a_\mu(\bx)}
-  
 \frac{\deltab^2 \BPsi }{\deltab A^a_\nu (\bx)\deltab A^a{}^\nu  (\bx)}   
  \nn \\ 
  && \nn \\
&& \hspace{-150pt} - \Tr
\, \Tr \left \{ 
 \frac{\delta^2 \BPsi}{\delta \Psi^T_\Sigma(\bx)\otimes\delta\Psi^T_\Sigma(\bx)} 
~\der_{A^a_\mu} \Psi_\Sigma (\bx)  
\otimes  \der_{A_a^\mu} \Psi_\Sigma  (\bx) \right\}
\label{reprod} \\
&& \nn \\
&& \hspace{-150pt} - 2\Tr \left\{
\frac{\delta \deltab \BPsi }{\delta \Psi^T_\Sigma(\bx)\deltab A^{a\nu}(\bx)}~\der_{A^a_\nu} \Psi_\Sigma(\bx)
\right \} . \nn 
\eeqa

The meaning of the 
replacement (\ref{722}) can be understood, e.g., with the lattice regularization in mind, when the regularized value of $\delta(\mathbf{0})$ can be expressed in terms of  the lattice spacing $a$ as $a^{-n+1}$. Then one can write the regularized version of (\ref{722}) 
also as $\frac1\ka\beta \mapsto a^{n-1}$. When $a$ is infinitesimal, the right-hand side of the latter expression becomes $\rd\bx$ and thus the 
substitution (\ref{722}) can be seen as the inverse of the quantization map (\ref{qmap}) underlying precanonical quantization. 
Moreover, the relation defining the delta function: 
$\int\! d\bx\, \delta(\bx) = 1$, can be interpreted as a definiton of the inverse of $d\bx$ which  
 is represented by $\frac{1}{\ka}\beta$ under the precanonical quantization,  eq. (\ref{qmap}). 
 Hence, the inverse of $d\bx$, a regularized delta function localized at $\bx = \mathbf{0}$, is 
 given by the inverse of $\frac{1}{\ka}\beta$:  $\beta\ka \mapsto \delta(\mathbf{0})$, which is exactly the singular limiting map in (\ref{722}). Since the map (\ref{qmap}), which replaces differential forms by the elements of Clifford algerba, is an essential part of precanonical quantization, its inverse in (\ref{722}), which will be shown to be a crucial element of transition from the precanonical quantization to the functional Schr\"odinger representation, can be seen as a partial dequantization.

\bigskip

The third term on the right-hand side of (\ref{reprod}) can be shown to be vanishing - 
under the replacement (\ref{722}) - by the following argument. 
Let us notice that  any new term $-U$ added to the pure Yang-Mills Lagrangian, 
which, e.g., may describe interactions with other fields,  
adds a term $+U$ to the DW Hamiltonian operator $\what{H}$ in (\ref{ym-nse}) 
(c.f. its derivation in \cite{my-ym1,my-ehrenfest}) 
and consequently, it will add the additional  term 
\beq \label{616}
\mathtt{Tr} \int\!d\bx\  \frac{\delta \BPsi}{\delta \Psi^T_\Sigma (\bx)} 
\frac{1}{\ka}\beta U (\bx)
\Psi_\Sigma (\bx)
\eeq 
on the right-hand side of (\ref{dtpsi}).  If the functional Schr\"odinger representation can be 
deduced from the precanonical description, this term should 
correspond to  the additional term 
$\int\!d\bx\, U(\bx) \BPsi $ in the canonical Hamiltonian operator acting on the 
Schr\"odinger wave functional in  (\ref{fseq}).  
 Because the term $U$ can be an arbitrary function 
of YM fields and external fields, this correspondence 
can be only possible if 
 $\mathtt{Tr} \left\{\frac{\delta \BPsi}{\delta \Psi^T_\Sigma (\bx)} \frac{1}{\ka}\beta \Psi_\Sigma (\bx) \right\}$ 
 in (\ref{616})  
 reproduces  $\BPsi$ 
 for any $\bx$, i.e. it is required that 
\beq \label{915}
\forall \bx: \quad \mathtt{Tr} \left\{\frac{\delta \BPsi}{\delta \Psi^T_\Sigma (\bx)} 
\frac{1}{\ka}\beta 
\Psi_\Sigma (\bx) \right\} 
\mapsto \BPsi~.  
\eeq
Hence, the dependence of the functional $\BPsi$ on $\Psi_\Sigma(\bx)$ 
is identical at all points $\bx$. 
This observation suggests that the functional $\BPsi$ 
 can be expressed 
in terms of a continuous product over all points $\bx$ 
of identical expressions including $\Psi_\Sigma (\bx)$.   
In this case, however,  
$\mathtt{Tr} \left\{\frac{\delta \BPsi}{\delta \Psi^T_\Sigma (\bx)} 
\Psi_\Sigma (\bx) \right\} 
= \delta(\mathbf{0}) \BPsi$. Therefore,  again, the map in (\ref{915}) 
  exists only 
  when $\frac{1}{\ka}\beta \delta(\mathbf{0}) \mapsto 1$ which is equivalent to the 
  substitution given by (\ref{722}).  
 Then, by acting by $\frac{\delta}{\delta \Psi_\Sigma(\bx)}$  
on both sides of (\ref{915}),  we conclude, similarly to eq. (3.15) in \cite{atmp2}, 
that the third term in (\ref{reprod}) vanishes 
 under the very same substitution (\ref{722}):  
\beq \label{vanish}
\Tr \, \Tr \left \{ 
 \frac{\delta^2 \BPsi}{\delta \Psi^T_\Sigma(\bx)\otimes\delta\Psi^T_\Sigma(\bx)} 
~\der_{A^a_\mu} \Psi_\Sigma (\bx)  
\otimes  \der_{A^{a\mu}} \Psi_\Sigma  (\bx) 
\right \} 
 \mapsto 0 . 
\eeq

Therefore, from (\ref{reprod}) and (\ref{vanish}) it follows that the terms $(IV)$ and $(V)$ in (\ref{dtpsi}) 
 under the substitution (\ref{722}) 
 are mapped to the following expression in variational derivatives of $\BPsi$: 
\begin{align}
\begin{split}\label{reprod1}
 \Tr\left \{\frac{\delta \BPsi }{\delta\Psi^T_\Sigma(\bx, t)} 
 \frac12 \beta\ka\, \der_{A^a_\mu}\der_{A_a^\mu} \Psi_\Sigma (\bx) \right \}
 \mapsto\ &
\frac12 \frac{\deltt{}^2 \BPsi }{\deltt\! A_a^\mu(\bx)\deltt\! A^a_\mu(\bx)}
-  
\frac12 \frac{\deltab^2 \BPsi }{\deltab A^a_\nu (\bx)\deltab A^a{}^\nu  (\bx)}   
  \\ 
  & \\ 
&\hspace{-5pt} - 
\Tr \left\{ \frac{\delta \deltab \BPsi }{\delta \Psi^T_\Sigma(\bx)\deltab A^{a\nu}  (\bx)} 
\der_{A^a_\nu} \Psi_\Sigma(\bx)
\right \} .  
\end{split}
\end{align} 

\bigskip 

Our next step is to consider the terms with $\der_{A_0^a}\Psi_\Sigma(\bx)$ in (\ref{dtpsi}). Namely,  
(\ref{term2+6}) and the last term in (\ref{reprod1}) corresponding to $\nu=0$. 
 Their total contribution to the 
functional derivative Schr\"odinger equation, which should not have explicit terms with 
$\Psi_\Sigma$, should be zero. Therefore, from (\ref{term2+6}) and the temporal part 
of the last term in (\ref{reprod1}), which is a part of the term (V) in (\ref{dtpsi}), we obtain 
\beq  \label{p24}
\frac{\ri}{2}\frac{\delta \BPsi}{\delta\Psi^T_\Sigma(\bx) }\gamma^{ij}F^a_{ij}(\bx) 
- \frac{\delta\deltab \BPsi}{\delta\Psi^T_\Sigma(\bx)\deltab A_0^a(\bx)} =0.
\eeq
Introducing the notation 
\beq \label{phi721}
\frac{\delta \BPsi}{\delta\Psi^T_\Sigma(\bx)} =: \BPhi (\bx)\,,
\eeq 
where $\BPhi(\bx)$ is 
an $\bx$-dependent Clifford-algebra-valued functional of $\Psi_\Sigma(\bx)$ and $A(\bx)$, 
eq. (\ref{p24}) takes the form 
\beq
\frac{\ri}{2}\BPhi (\bx)\gamma^{ij}F^a_{ij}(\bx) - \frac{\deltab\BPhi(\bx)}{\deltab A_0^a(\bx)}
=0. 
\eeq 
  The solution $\BPhi (\bx)$ can be written 
in the form which is implying the substitution (\ref{722}) again: 				 
\beq \label{bphi}
\BPhi (\bx) = \mathbf{\Xi} [\Psi_\Sigma (\bx); \check{\bx}] 
e^{\frac{\ri}{2\ka} \beta\gamma^{ij}A_0^a (\bx) F^a_{ij}(\bx)} 
{\big|_{\beta\ka \mapsto \delta({\mathbf 0})}}  \,,
\eeq
where $\mathbf{\Xi} [\Psi_\Sigma (\bx); \check{\bx}]$ is a functional of 
$\Psi_\Sigma(\bx)$ 
 on the punctured space  $\mathbb{R}^{n-1}\backslash \{\bx\}$  
 (c.f.  eq.~(3.21) in \cite{atmp2}) 
which plays here a role of the integration constant. 
 Then the solution of (\ref{phi721}) takes the form 
  (up to a normalization factor which will also include $\ka$)  
\beq \label{p26}
\BPsi \sim  \Tr \Big\{
\mathbf{\Xi} [\Psi_\Sigma (\bx); \check{\bx}]  e^{\frac{\ri}{2\ka} \beta\gamma^{ij}A_0^a (\bx) F^a_{ij}(\bx)} 
{\frac{\beta}{\ka}}
\Psi_\Sigma(A_\mu^a(\bx)) 
\Big\}{\raisebox{-3pt}{$\big|_{\beta\ka \mapsto \delta({\mathbf 0})}$}}
, 
\eeq
 where the multiplier ${\frac{\beta}{\ka}}$ enters to the left from 
$\Psi_\Sigma$ in order  to compensate ${\delta({\mathbf 0})}$ which will appear from the variation of 
$\Tr\{\BPhi(\bx)\Psi_\Sigma(\bx)\}$ with respect to $\Psi_\Sigma(\bx)$. 
Note that the solution in this form is consistent with (\ref{915}) which was already established by means of an independent reasoning.

Since the formula (\ref{p26}) should be valid for any $\bx$, 
we conclude that the wave functional $\BPsi$ has the structure of  
a continuous product  over $\bx$: 
\beq 
\BPsi \sim  \Tr 
\prod_\bx \left \{ e^{\frac{\ri}{2\ka} \beta\gamma^{ij}A_0^a (\bx) F^a_{ij}(\bx)} 
 {\frac{\beta}{\ka}}
\Psi_\Sigma(A_\mu^a(\bx)) 
\right\}\!{\raisebox{-3pt}{$\big|_{\beta\ka \mapsto \delta({\mathbf 0})}$}}
.  
\eeq
This rather symbolic expression can be understood, up to a normalization factor,  as the 
multiple Volterra product integral over $\bx$ \cite{prodint} denoted 
as $\prodi_\bx f(\bx)^{\rd \bx}$:   
\beq \label{p27}
\BPsi \sim \Tr\, 
\underset{\!\!\bx}{\scalebox{1.5}{$\displaystyle \prodi$}}
\left \{ e^{\frac{\ri}{2\ka}\beta \gamma^{ij}A_0^a (\bx) F^a_{ij}(\bx)}
 {\frac{\beta}{\ka}}
   \Psi_\Sigma(A_\mu^a(\bx))\right\}\!{\raisebox{-3pt}{$\big|_{\frac{1}{\ka}\beta \mapsto \rd \bx}$}} .
\eeq 
  
Let us note here that the 
 multiple product integral is very little known and explored, while the one-dimensional product integral is well known under different names such as the Peano series or ``path (or time) ordered exponential" used in QFT  (see also \cite{mansouri}). A multidimensional generalization of the ``path ordering" is discussed in 
\cite{prodint} and it is probably not unique. Nevertheless, the expression (\ref{p27}) already 
serves the purpose of the paper as it shows that the description in terms of the 
Schr\"odinger wave functional, which is known to require additional constructs, such as regularization, in order to be usable within the existing QFT (see e.g. \cite{jackiw,luscher}), emerges from the well defined precanonical description only in a limiting case when both the 
 substitution  
$\frac{1}{\ka}\beta \mapsto  \rd \bx$    and the multiple product integral itself have to be understood as the continuum limits of certain discretized or otherwise regularized expressions, 
which also require additional constructs in order to be mathematically correctly  defined.  

\bigskip

Next, after integrating  by parts and using 
 the constraint $\hat{\pi}{}_a^{(i0)}\Psi_\Sigma \approx 0$,  
 the term $(III)$ in (\ref{dtpsi}) transforms to 
\beq \label{term3}
\mathit{III}:\quad 
\ri A_0^a(\bx)\  
 \der_i \Tr  \left( \BPhi(\bx)  \der_{A_i^a}\Psi_\Sigma(\bx) \right) 
. 
 \quad 
\eeq
Using  the constraint 
$\what{\pi}{}_a^{(i0)}\Psi_\Sigma \approx 0$ 
in $(VIII)$, 
the of sum   $(VII)$ and $(VIII)$ yields  
 \beq \label{term7+8}
\mathit{VII+VIII}:\quad   
\ri g C^a_{bc} A^b_0(\bx) A^c_i(\bx)
\Tr \left\{ \BPhi(\bx)
 \der_{A^a_i}\Psi_{\Sigma}(\bx) \right\}  . 
\eeq 
Therefore, the terms  $(III), (VII)$ and $(VIII)$ together 
lead to 
\beq \label{3+7+8}
\mathit{III+VII+VIII}:\quad 
\ri A_0^b(\bx) D^{\,\, a}_{ib}
\Tr \left\{ \BPhi(\bx)
\der_{A_i^a}\Psi_\Sigma(\bx) \right\} \,, 
\eeq 
where  $D^{\,\, a}_{ib} := \delta_b^a \der_i + g C^a_{bc} A^c_i(\bx) $ 
denotes the covariant derivative. 
  By remembering the expression of the first total variational derivative in 
(\ref{delta-bpsi1}), we see that 
the   terms $(III), (VII)$ and $(VIII)$ in (\ref{dtpsi})   yield  
 \beq \label{gauss0}
 \ri \int\!d\bx\ A^b_0(\bx) \Big( 
 \delta_b^a \der_i + g C^a_{bc} A^c_i(\bx)\Big) 
 \left(
 \frac{\deltt \BPsi}{\deltt\! A^a_i(\bx)} 
 - \frac{\deltab \BPsi}{\deltab A^a_i(\bx)}\right), 
 \eeq
 that resembles  the standard appearance of the quantum Gau\ss\ constraint 
in the canonical Hamiltonian operator of YM theory (see e.g. \cite{huang}). 

\bigskip 

Now, let us consider the total derivative in term $(I)$ 
in (\ref{dtpsi}). By integration by parts 
and using the cyclic permutation under the trace, 
we obtain 
\beq \label{dxpsi}
  - \int\!d\bx\ \mathtt{Tr} \left\{ \frac{\delta \BPsi}{\delta \Psi^T_\Sigma (\bx)}
 \alpha^i \frac{d}{dx^i} \Psi_\Sigma(\bx) \right\} 
 =  \int\!d\bx\ \mathtt{Tr} 
 \left\{ \frac{d}{dx^i}\BPhi (\bx) \alpha^i \Psi_\Sigma(\bx)\right\}  \,.
 \eeq
Using  (\ref{bphi}),  on the right-hand side of (\ref{dxpsi}) we have 
\beq
\int\!d\bx\ \mathtt{Tr} \left\{ 
\BPhi(\bx) 
\frac{d}{dx^i}\left( -\frac{\ri}{2\ka} \beta\gamma^{kl}A_0^a (\bx) F^a_{kl}(\bx) \right) 
\alpha^i 
\Psi_\Sigma(\bx)  
\right\}\raisebox{-4pt}{$\Big|_{\beta\ka \mapsto \delta({\mathbf 0})}$} \,.
\eeq
By the cyclicity of trace it equals to 
\begin{align}
&\int\!d\bx\ \mathtt{Tr} \left\{ 
\Psi_\Sigma(\bx) \BPhi(\bx) 
\frac{d}{dx^i}\left( -\frac{\ri}{2\ka} \beta\gamma^{kl}A_0^a (\bx) F^a_{kl}(\bx) \alpha^i \right) 
\right\}\raisebox{-4pt}{$\Big|_{\beta\ka \mapsto \delta({\mathbf 0})}$} \nn \\ 
 &  \label{p748}\\
&  = \mathtt{Tr} \left\{ \beta ||\BPsi||\int\!d\bx\
\frac{d}{dx^i}\left( -\frac{\ri}{2} \gamma^{kl}A_0^a (\bx) F^a_{kl}(\bx) \gamma^i \right) 
\right\} \raisebox{-4pt}{$\Big|_{\beta\ka \mapsto \delta({\mathbf 0})}$} = 0 \,, \nn
\end{align}
where $||\BPsi|| := \frac1\ka\beta\Psi_\Sigma(\bx) \BPhi(\bx) $ is the matrix-valued functional whose trace is $\BPsi$ (c.f. (\ref{bphi}), (\ref{p26})), hence $\bx$-independent. The integral on the right-hand side of (\ref{p748}) vanishes as its integrand is a total divergence. Therefore, we have shown that the 
term $(I)$  in (\ref{dtpsi}) does not contribute to the functional derivative equation for $\BPsi$  
(again, under the limiting map (\ref{722})): 
\beq
\mathit{I}: \quad \int\!d\bx\ \mathtt{Tr} \left\{ \frac{\delta \BPsi}{\delta \Psi^T_\Sigma (\bx)}
 \alpha^i \frac{d}{dx^i} \Psi_\Sigma(\bx) \right\} 
\mapsto 0.
\eeq

\bigskip 

 Our next step is to consider  the $\der_{A^a_0}\der_{A_a^0} \Psi_\Sigma (\bx)$ term 
$(V)$ in (\ref{dtpsi}). 
  To this end, let us recall that the behavior of $\Psi_\Sigma(\bx)$ is determined by the  precanonical Schr\"odinger equation restricted to $\Sigma$, eq.~(\ref{psisigmaeqn}). Our  analysis above has shown that the total derivative term 
in (\ref{psisigmaeqn}) has no contribution to the functional Schr\"odinger equation and that some other terms in (\ref{dtpsi}) play together to produce the Gau\ss{} constraint (c.f  (\ref{gauss0})). 
Hence  the part of the time evolution of $\Psi_\Sigma(\bx)$ which is relevant for the time evolution 
of the Schr\"odinger wave functional $\BPsi$ is controlled by the remaining terms 
inside the round brackets in  (\ref{dtpsi})  (c.f. (\ref{term2+6})): 
\beq \label{nsesigmaga}
\ri \der_t \Psi_{\Sigma} \approx \beta \left( \frac{\ri}{2}\beta\gamma^{ij}F^a_{ij}(\bx) 
\der_{A^a_0}  + \frac\ka2 \der_{A_0^aA_0^a} -  \frac\ka2 \der_{A_i^aA_i^a} \right)
\Psi_\Sigma .
\eeq
This equation can be seen as the restriction of precanonical Schr\"odinger equation (\ref{ym-nse}) to $\Sigma$\, 
{\sl and}{\,}  the subspace of constraints given by (\ref{qconstr2}) and the initial value (Gau\ss) constraint. 
By writing the r.h.s. of (\ref{nsesigmaga}) in the form of a ``magnetic Schr\"odinger operator"  
with a constant Clifford-valued ``magnetic potential" in the space of temporal components $A_0^a$, i.e. 
\beq \label{nse-r}
\ri \der_t \Psi_{\Sigma} = \beta \left( 
- \frac12 \left( \ri \sqrt{\ka}\der_{A_0^a} - \frac{1}{2\sqrt{\ka}}\beta\gamma^{ij}F^a_{ij}(\bx) \right)^2  
- \frac{1}{4\ka}  F^a_{ij}(\bx)F^a{}^{ij}(\bx) -  \frac\ka2 \der_{A_i^aA_i^a} \right)
\Psi_\Sigma , 
\eeq  
 we conclude that the dependence of $\Psi_\Sigma$ on $A_0^a (\bx)$ can be absorbed in the phase factor:     
\beq \label{psia0}
\Psi_\Sigma (A_0^a(\bx),A^a_i(\bx), t) = (1+\beta) 
e^{-\frac{\ri}{2\ka} \beta \gamma^{ij}A^a_0(\bx)F^a_{ij}(\bx)}\, \Phi_\Sigma (A^a_i(\bx),t) ,
\eeq 
where the projecting factor $(1+\beta)$ is present because of the 
identity $\beta(1+\beta)= 1+\beta$. 
Then 
  \begin{align} \label{14ff}
  \frac{1}{2}\beta \ka\, \der_{A_0^aA_0^a}
  \Psi_\Sigma(\bx){} = 
   \frac{1}{2}\beta \ka   \left( \frac{-\ri\beta}{2\ka}\gamma^{ij} F^a_{ij}\right)^2\Psi_\Sigma (\bx)  =  \frac{\beta}{4\ka } \ F^a_{ij}F_a^{ij} \Psi_\Sigma (\bx), 
  \end{align} 
  where the identity $(\gamma^{ij} F_{ij})^2 = -2F_{ij}F^{ij}$ is used. 
Therefore, by taking into account (\ref{915}), 
the term $(V)$ in (\ref{dtpsi}) under the limiting map (\ref{722})  yields: 
\beq \label{238}
 V: \quad \Tr \int\! \rd\bx \left\{\BPhi(\bx)   \frac{1}{2}\beta \ka\, \der_{A_0^aA_0^a}
  \Psi_\Sigma(\bx) \right\} \mapsto \frac{1}{4}  F_{ij}F^{ij} \BPsi .
\eeq 
Thus we see that the right hand side of (\ref{238}) correctly reproduces the magnetic energy term in (\ref{fseq}). 

\bigskip 

Further, from (\ref{915}) and (\ref{psia0}) we obtain, up to a normalization factor, 
\beq  \label{249}
\forall \bx: \quad 
\BPsi \sim \Tr 
 \left\{ \BPhi (\bx) \frac{1}{\ka} (1+\beta) e^{-\frac{\ri}{2\ka} \beta \gamma^{ij}A^a_0(\bx)F^a_{ij}(\bx)}\, \Phi_\Sigma (A^a_i(\bx),t)  \right\} \raisebox{-4pt}{$\Big|_{\beta\ka \mapsto \delta({\mathbf 0})}$} \,. 
\eeq 
By comparison with the form of  $\BPhi (\bx)$ obtained in (\ref{bphi}), we conclude that the 
phase factors with $A_0^a$ in $\BPhi(\bx)$ and  $\Psi_\Sigma (\bx)$ 
actually cancel each other at each point $\bx$, so that 
\beq
\forall \bx: \quad 
\BPsi \sim \Tr 
 \left\{ \mathbf{\Xi} (\check{\bx}) \frac{1}{\ka} (1+\beta) \Phi_\Sigma (A^a_i(\bx),t) 
  \right\}\raisebox{-4pt}{$\Big|_{\beta\ka \mapsto \delta({\mathbf 0})}$} \,.
\eeq 
This again suggests that (up to a normalization factor which may include $\ka$) 
the functional $\BPsi$ is expressed as the product integral of wave functions  
$(1+\beta) \Phi_\Sigma (\bx)$: 
\beq
\BPsi \sim \Tr\ \underset{\!\!\bx}{\scalebox{1.5}{$\displaystyle \prodi$}}
\Big\{  (1+\beta) \Phi_\Sigma (A^a_i(\bx),t)  
\Big\}\raisebox{-4pt}{$\Big|_{{\frac{\beta}{\ka}}\mapsto\rd\bx}$} \,.
\eeq
From this expression it is obvious that the partial variational derivatives of 
$\BPsi[\Phi_\Sigma (A^a_i(\bx),t)]$ 
with respect to $A_i^a(\bx)$ are actually vanishing: 
\beq 
\frac{\deltab\BPsi}{\deltab A^a_i(\bx)}=0 .
\eeq 
 Then
 the term $(IV)$ in (\ref{dtpsi}) 
 produces just the second variational derivative of $\BPsi$ (c.f. (\ref{reprod1})): 
 \beq
 IV:\quad  - \Tr \int\! \rd\bx \left\{\BPhi(\bx)   \frac{1}{2}\beta \ka\, \der_{A_i^aA_i^a}
  \Psi_\Sigma(\bx) \right\} 
 \mapsto
 - \frac{1}{2} \frac{\delta^2\BPsi}{\delta A_i^a(\bx){}^2},    
 \eeq
 and the last term in (\ref{gauss0}) is vanishing, so that the terms $(III), (VII)$ and 
 $(VIII)$ exactly reproduce the usual Gau\ss\, constraint term in the canonical Hamiltonian operator: 
 \beq \label{gauss1}
 III + VII + VIII: \quad 
 \ri \int\!d\bx\ A^b_0(\bx) 
 D_i{}^a_b 
 \frac{\delta \BPsi}{\delta\! A^a_i(\bx)} \,.
 \eeq


Summarizing the results of the above consideration of (\ref{dtpsi}), we have shown that 
\begin{itemize}
\item the terms $(III), (VII)$ and $(VIII)$ together with the antisymmetry constraint 
(\ref{qconstr2})  reproduce the Gau\ss\ constraint term in the canonical Hamiltonian operator, 
\item the term $(IV)$ reproduces the second  variational derivative term in the canonical Hamiltonian operator in the limiting case $\beta\ka \mapsto \delta({\mathbf{0}})$, 
\item the terms $(II), (V)$ and $(VI)$ together with the antisymmetry constraint 
(\ref{qconstr2})  allow us to obtain a product integral formula relating the 
Schr\"odinger wave functional and 
the precanonical wave function, and to reproduce the correct 
magnetic contribution $\frac14 F_{ij}^aF^{ij}_a$ to the Yang-Mills Hamiltonian operator 
 in the temporal gauge,\footnote{Note that in our previous paper \cite{my-ym1}, where the temporal gauge was imposed by hand, we actually obtained a wrong result $\frac18 F^2$ for the magnetic energy, see the erroneous transition from (4.17) and (4.20) to eq. (4.21).} 
\item the term $(I)$ does not contribute to the Hamiltonian operator (assuming  
 the precanonical wave function is vanishing at the spatial infinity). 
\end{itemize} 

By combining those results we conclude that, in the limiting case (\ref{722}), 
 we obtain from the precanonical Schr\"odinger equation (\ref{ym-nse}) and precanonical quantum constraints  (\ref{qconstr}) 
 the  following equation on the functional $\BPsi$: 
 \beq
 \ri\der_t\BPsi = \int\! \rd\bx  \left( - \frac{1}{2} \frac{\delta^2}{\delta A_i^a(\bx){}^2} 
 + \frac{1}{4}  F_{ij}(\bx) F^{ij}(\bx) + \ri A_0^a (\bx) D_i{}_a^b \frac{\delta}{\delta A_i^b(\bx){}} 
 \right) \BPsi . 
 \eeq 
 The temporal components $A_0^a$ appear here as the Lagrange multipliers 
 which fix the quantum version of the Gau\ss\, constraint (\ref{gauss}). 
 The rest of the equation is just the canonical Schr\"odinger equation for 
 quantum Yang-Mills field (\ref{fseq}) which one usually derives using the canonical quantization in the temporal gauge (see e.g. \cite{hatfield,huang}).

Thus, in the limiting case when $\beta\ka$ goes to (a regularized value of) $\delta{\bf(0)}$ 
 (i.e. essentially to the UV cutoff of the momentum space volume),
 we have derived from the precanonical Schr\"odinger equation (\ref{ym-nse}) and the quantum antisymmetry constraints (\ref{qconstr}) the canonical 
functional derivative Schr\"odinger equation (\ref{fseq}) in the temporal gauge, 
 the Gau\ss\ constraint (\ref{gauss}), and the explicit product integral formula (\ref{p27}) 
relating the Schr\"odinger wave functional with the Clifford-valued 
precanonical wave function.


\bigskip

\section{Conclusion} 

We have shown in our previous papers that 
the approach of precanonical quantization leads to the description 
of quantum pure Yang-Mills theory in terms of the Clifford-algebra-valued 
precanonical wave functions on the space of Yang-Mills field components 
$A^a_\mu$ and space-time coordinates $x^\mu$.  
The precanonical wave function satisfies the 
analogue of the Schr\"odinger equation defined on the aforementioned bundle, 
eq. (\ref{ym-nse}), 
which treats all space-time dimensions on  equal footing. 
 The expressions of precanonical quantum operators 
typically contain an ultraviolet parameter $\varkappa$ of the dimension of an inverse spatial volume.  

This description contrasts with the more familiar description of quantum YM fields 
in terms of the Schr\"odinger wave functionals of field configurations $A^a_\mu (\bx)$ 
at fixed moments of time $t$ which is derived from canonical quantization and implies a distinction between the space variable $\bx$ and the time variable $t$ 
\cite{hatfield,huang}. 

In this paper,  we  investigated a  connection between the description of quantum YM theory 
based on precanonical quantization and  the standard description  
 in the functional Schr\"odinger representation.   
 We demonstrated that the latter can be derived from the former in the 
limiting case of an infinitesimal value of $1/ \varkappa$ 
when the Clifford algebra element $\gamma_0 / \varkappa $ can be mapped 
to the differential form representing an infinitesimal volume element $\mathrm{d} \bx$.  In this singular limiting case,  
we were able to derive both the standard functional derivate Schr\"odinger equation for 
the quantum YM field and the Gauss law constraint from the precanonical 
Schr\"odinger equation, and also to obtain an expression of the 
Schr\"odinger wave functional of quantum YM theory in terms of 
a multiple product integral of precanonical wave functions restricted to a field configuration 
 $A_\mu^a = A_\mu^a (\bx)$. 
 
Our results suggest that the standard functional Schr\"odinger representation of 
quantum field theory of Yang-Mills fields 
appears from the precanonical formulation  as a singular limiting case. 
 While the  
former, in order to be a well-defined theory at least on the physical level of rigour, 
is known to require an ad hoc regularization (e.g. a point split in the second variational 
derivative in the functional derivative Schr\"odinger equation (\ref{fseq})), which typically 
introduces a UV cutoff scale $\Lambda$ as a necessary additional element of the theory 
removed by a subsequent renormalization,  
 the precanonical formulation is 
``already regularized" 
by introducing the ultraviolet scale $\varkappa$ as an inherent element of the precanonical quantization procedure. 

One should notice that the introduction of the ultraviolet scale $\ka$ in precanonical quantization 
does not introduce a grainy structure of space-time at ultra-small scales or violate the 
relativistic space-time at the corresponding high energies. In the case of quantum mechanics, 
which corresponds to one-dimensional space-time, $\ka$ is dimensionless and the Clifford-algebra-valued precanonical wave function is complex valued, hence a usual quantum mechanics is reproduced. 
The question still remains if $\varkappa$ is a universal fundamental scale or it is 
an auxiliary 
element of precanonical quantization of fields 
which should be removed from the physical results  by a procedure similar 
to the usual renormalization. 

Quite unexpectedly, our recent estimations of the scale of $\ka$ based on the estimation of the cosmological constant in precanonical quantum gravity \cite{qg} and the estimation of the YM mass gap \cite{my-massgap} have consistently pointed to the possible subnuclear scale of $\ka$, which contradicts a more plausible naive expectation that the scale of $\ka$ is Planckian. 
The physical meaning of these results is still to be explored. 

 \paragraph*{Acknowlegdements:} 
I gratefully appreciate the hospitality of the School of Physics and Astronomy 
of the University of St Andrews, Scotland, and 24/7 availability of its facilities for research.  


\end{document}